# INTRODUCTION TO BULK METALLIC GLASS COMPOSITE AND ITS RECENT APPLICATIONS

*Shijing Lu*

Material Scienc and Engineering Department, North Carolina State Universtity
slu3@ncsu.edu

**Abstract**

Bulk metallic glass (BMG) materials are hot topics in recent years, not to mention BMG matrix composites, which further improve the magnetic and mechanical properties of BMG materials. BMG and BMG matrix materials are fast developing and promising materials in modern industry due to their extraordinary properties such as high strength, low density, excellent resistibility to high temperature and corrosion. In this paper, I reviewed processing and application of several recently developed BMG and BMG matrix materials.

**Key-words:** Bulk metallic glass; Composites; in-*suit*

## 1  General idea of glass matrix composites

Nowadays glass materials are seen everywhere, from ceramic house wares to ceramic materials in all buildings. This is because glass materials are strong in compression, can withstand most chemical erosion and high temperature. However glass materials are brittle, so that they cannot withstand high shear stress. One of the most efficient solutions to solve the problem of brittleness is to embed other high strength phases in the ceramic matrix, which is often know as glass matrix composites. The embedded phases, which is also called strengthening phases, provides occurrence of processes that can stop propagation of failures and therefore strengthen the materials on the whole.

## 2  Processing of metallic glass

A dominate factor in formation and processing of metallic glass is cooling rate for glass formation. The first sample of metallic glass of $Au_{75}Si_{25}$ made by Duwez at Caltech in 1960[1.1], was by rapidly quenching metallic liquid at a very high cooling rate ($10^5 \sim 10^6 \, K/s$). However at high cooling rate, the amorphous alloy geometry was limited to small particles or thin sheets, which are unlikely to have wide applications. For a long time, researchers were devoted to finding ways of formatting bulk metallic glass. The breakthrough was contributed by Turnbull and his coworkers [1]. Turnbull predicted that the glass forming ability (GFA) is determined by a ratio, known as reduced glass transition temperature $T_{rg} = T_g/T_m$, where $T_g$ is the glass transition temperature, and $T_m$ is the melting temperature of the metallic liquid. According to Turnbull's theory, a liquid with $T_{rg} = 2/3$ has the greatest GFA and can form metallic glass at a very low cooling rate.

Promoted by more and more comprehensive fundamental understanding of metallic glass of materials, the first reported bulk metallic glass of Pd-Cu-Si alloy was prepared by Chen in 1972 [2] at a relatively low cooling rate of $10^3 K/s$. Later on, many bulk metallic glass materials are manufactured and their dimension varies from a few micrometers to several centimeters [3]. The cooling rate reaches a level as low as $\sim 1 K/s$ [4].

## 3  Properties of metallic glass

It is some excellent mechanical and chemical properties that makes metallic glass materials a hot topic in material science and engineering nowadays. In general metallic glass differs from normal crystalline alloys for following two aspects.

### 3.1  High tensile strength and low Young's moduli.

As demonstrated in Fig. 1, both tensile strength and Vickers hardness are roughly linearly dependent on Young's moduli. Compared to normal crystalline alloy which was plotted as hollow symbols in the figure, metallic glass has approximately 40% lower value in Young's moduli. Or in another saying, metallic glass materials have higher tensile strength and Vickers hardness.

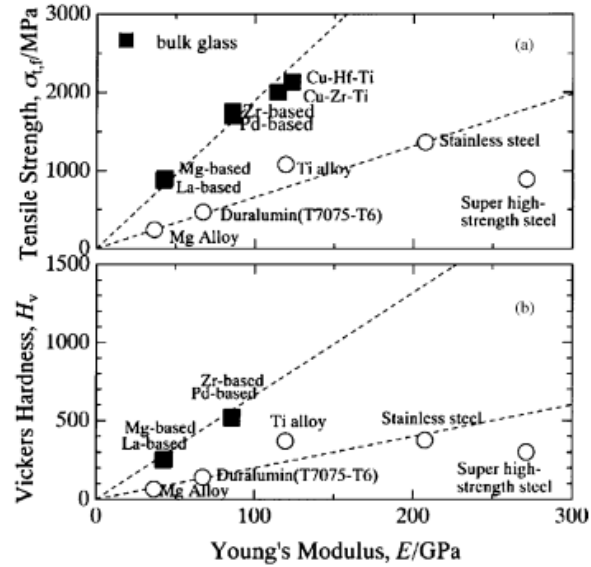

**Fig. 1** The dependence of Vickers Hardness and tensile failure strength on Young's moduli for metallic glass materials (Copyright (2002) by The Japan Institute of Metals.)

## 3.2  Good soft magnetism

Properties of soft magnetism are often seen in Fe and Co based BMG. For example, Permeability of a Co based BMG which has a ring shape of 1mm thickness, 5mm inner diameter, 10 mm out diameter, reaches a maximum value of 500,000 and minimum value of $0.26\ A\,m^{-1}$ [5]. Fig. 2 demonstrates another example of Fe based ring shape BMG. It can be seen clears from the figure that the annealing treatment of ring shape alloy cause an increase of initial permeability from 27,000 to 110,000 and a decrease of coercive force from $3.7\ Am^{-1}$ to $2.2\ Am^{-1}$ [6].

Since the excellent soft magnetism are due to highly homogeneous magnetic domain structure align the circumferential direction in the atomic scale, the soft magnetic properties of a amorphous thin film are structure sensitive and can easily be controlled by synthesizing process.

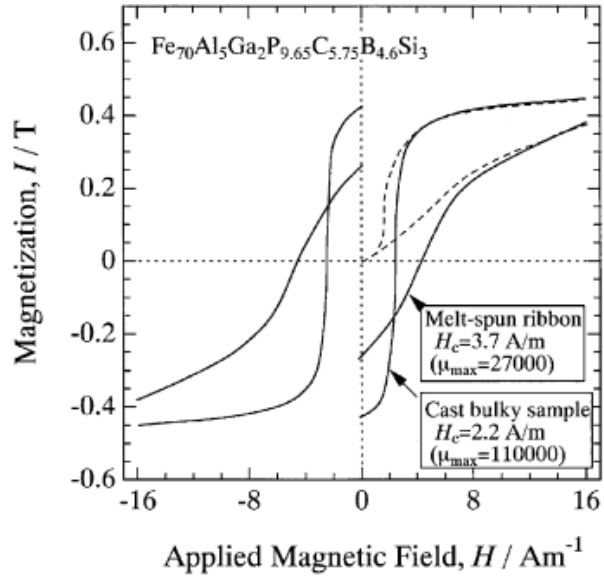

**Fig. 2** A comparison of I-H hysteresis loop of the Fe-based ring shape BMG and similar shape crystalline alloy. (Copyright (2000) by The Japan Institute of Metals.)

## 4  Processing and properties of BMG composites

BMG matrix composites are natural materials in the sense that a second phase are usually inevitable in the process of synthesizing BMG materials. For example in preparing Vitreloy I ($Zr_{41.2}Ti_{31.8}Cu_{12.5}Ni_{10}Be_{22.5}$) BMG, a composite composed of a ductile phase and embedding BMG particles are obtained [7].

Fig. 3 summarizes the processing and the microstructures of the above mentioned BMG composites. This composite, compared to single phase BMG materials, demonstrates strongly improved Charpy impact toughness and ductility. The three stages process shown in the Fig. 3 are genetic ways of preparing in suit BMG composites.

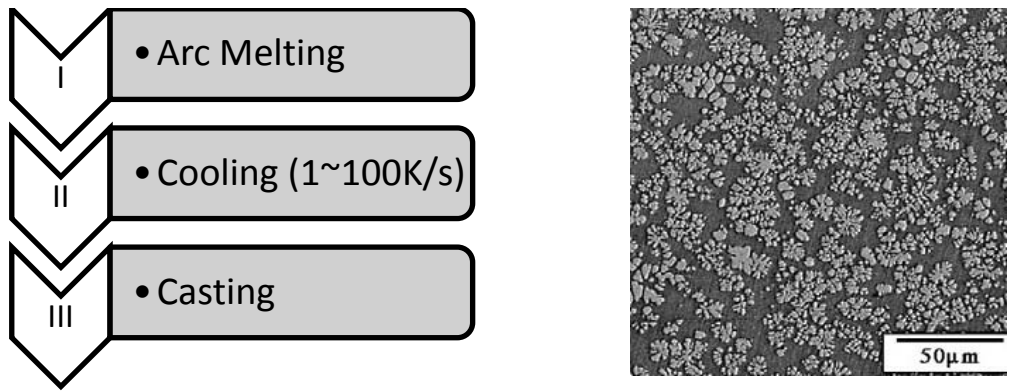

Fig. 3 Processing and final structure of Vitreloy base BMG composites. The process schetched on the left is a general ways of preparing BMG matrix composites.

In *suit* BMG composites is widely used and one of the most important ways of increasing plasticity or malleability of BMG materials. In Table 1 lists several examples of in *suit* BMG composites. Corresponding microstructures are summarized in Fig. 4.

**Table 1** Recently reported in suit BMG composites that designed to enhance plasticity of BMG materials

| Base Metal | BMG | Second Phase |
|---|---|---|
| Zr | $Zr_{41.2}Ti_{31.8}Cu_{12.5}Ni_{10}Be_{22.5}$ (7) | In *suit* ductile $\beta$ phase |
| | $Zr_{55.0}Cu_{29.0}Ni_{8.0}Al_{8.0}$ (8) | $(Zr,Ni,Al)_2(Cu,Ni,Al)$ intermetallic spheres |
| | $Zr_{31-34}Ti_{17-22}Nb_{1-2}Cu_{9-13}Be_{31-38}$ [4.7] (9) | In *suit* dendrite crystalline |
| Mg | $Mg_{81}Cu_{9.3}Zn_{4.7}Y_5$ [5] (10) | Mg + flake-shaped precipitates |
| | $(Mg_{0.65}Cu_{0.075}Ni_{0.075}Zn_{0.05}Ag_{0.05}Y_{0.1})_{1002x}Fe_x$ (x = 59 and 13) [4.17] (11) | In suit ductile phase |
| La | $La_{86-y}Al_{14}(Cu,Ni)_y$ [4.18] (12) (y = 1–24) | In *suit* ductile phase |
| Cu | $(Cu_{0.5}Hf_{0.35}Ti_{0.1}Ag_{0.05})_{100-x}Ta_x$ [4.19] (13) (x = 0, 1, 3, 5, 8, 12) | Ta-rich dendrites precipitates |
| | $(Cu_{0.6}Hf_{0.25}Ti_{0.15})_{90}Nb_{10}$ [4.20] (14) | In suit ductile phase |
| Pd | $Pd_{60-x}Ni_{10}Cu_{30}P_x$ [4.21] (15) x = 10 − 20 | F.C.C α-Pd crystalline phase |

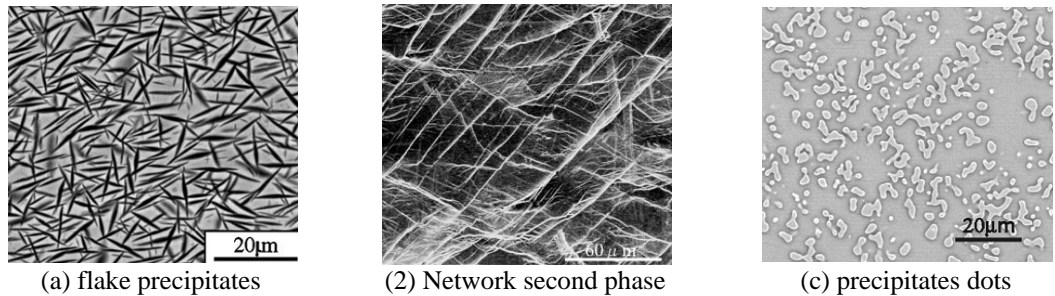

(a) flake precipitates    (2) Network second phase    (c) precipitates dots

**Fig. 4** Three commonly seen microstructures of in suit BMG composites. (Copy rights by 2006 Acta Materialia Inc)

## 5  Applications of BMG and its composites

Due to its unique and unconventional properties, BMG materials are good candidates for many applications. Reference (16) reviewed many commercial applications and potential applications reported in the literature. Here we briefly summarized, on the other hand, some of widely cited USA patents that are related to BMG materials or BMG matrix composites. The summary is presented Table 2.

**Table 2 Application fields of BMG or BMG composites in USA patent database.**

| BMG or BMG composites | Applications | Properties |
|---|---|---|
| Graphite Particles enforced composites in Zr based BMG matrix (17) | Joints, frictional bearings or springs. | Up to double the fracture strength and four times elasticity of their crystalline alloy conterparts. |
| Bulk solder materials from alloys processing deep eutectics with asymmetric liquidous slop. (18) | Printed electric circuit or semiconductor devices | High strength, high elasticity; physically electrically or thermally couple one feature to another. |
| Fe based BMG materials | soft magnetic materials for common mode choke coils | Soft magnetism |
| Ductile metal enforced BMG composites (19) | Spring, mechanical components and sporting goods | High plastic deformation |
| ZrAl(Fe, Co) or ZrAl(Pd, Au, Pt) type BMG materials (20) | solder bump for communication between an integrated circuit device | Low thermal stress under thermal processing, low modules that resists |

| | and external structures | cracking during shock and dynamic loading. |
|---|---|---|
| All BMG materials that are not specifically designed for good soft or hard ferromagnetic properties. (21) | Medial, surg, MRI-compatible medical instruments. | Biological compatible, low susceptibility effects, high elastic limit of about 2% as compared to that of a typical metal, namely about 1%. |